\documentstyle[12pt,epsf]{article}
\newcounter{fixy}
\begin{document}
\newenvironment{fixy}[1]{\setcounter{figure}{#1}}
{\addtocounter{fixy}{1}}
\renewcommand{\thefixy}{\arabic{fixy}}
\renewcommand{\thefigure}{\thefixy\alph{figure}}
\setcounter{fixy}{1}

\title{The Born-Infeld Sphaleron}
\author{{\large Yves Brihaye} \\
{\small Facult\'e des Sciences, Universit\'e de Mons-Hainaut, }\\
{\small B-7000 Mons, Belgium }\\
{ } \\
   {\large Betti Hartmann}\\
{\small Department of Mathematical Sciences ,
University of Durham}\\
{\small Durham DH1 \ 3LE , United Kingdom}}

\date{\today}
%%%%%%\begin{titlepage}
\maketitle
\thispagestyle{empty}
\begin{abstract}
We study the SU(2) electroweak model in which the standard Yang-Mills
coupling is supplemented by a Born-Infeld term. The deformation of the
sphaleron and bisphaleron solutions due to the Born-Infeld term 
is investigated and new
branches of solutions are exhibited. Especially, we find a new branch
of solutions connecting the Born-Infeld sphaleron to the
first solution of the Kerner-Gal'tsov series.
\end{abstract}
%%%\vfill
%%%%%%%%\end{titlepage}
%%%\newpage
\section{Introduction}
While in the pure Yang-Mills (YM) theory no particle-like, finite energy
solutions are possible, they might exist as soon as scale
invariance breaking terms
appear in the Lagrangian. These can be of different types.
The possibility discovered first is the coupling of the YM system
to a scalar Higgs field. This breaks the scale invariance and
particle-like solutions exist: a) monopoles \cite{hp},
when the Higgs field is given in the adjoint representation of $SU(2)$ and
b) sphalerons \cite{km}, when the Higgs field is given in the
fundamental representation of $SU(2)$. While these solutions were studied
extensively since their discovery, the coupling of gravity to YM theories
was rather neglected. It was believed that gravity was to weak to have important
effects on these solutions. The interest only rose when Bartnik and McKinnon
\cite{bm} found a particle-like, finite energy solution of the coupled
Einstein-YM system.

Recently, it was discovered that superstring theory induces important modifications
for the standard quadratic YM action.
Likely the relevant effective action is of the Born-Infeld type \cite{tse}
and, naturally, a scale invariance breaking term is involved.
This was exploited to construct
classical "glueballs" in the SU(2) Born-Infeld theory \cite{kg}.

Considering the classical equations of non-abelian
field theories coupled to various other fields leads -
depending on the coupling constants - in general to a rich pattern of
solutions involving several branches related
by bifurcations and/or connected by catastrophe-like cusps
\cite{bk,yaffe,bks}.

In this paper we investigate the effect of a Born-Infeld term
on the sphaleron and bisphaleron solutions arising in the SU(2)
part of the electroweak model
and we find that these solutions are smoothly deformed up to
a critical value of the Born-Infeld parameter $\beta_{BI}^2$. From there,
another branch of solution exists which connects the sphaleron
branch to the first solution of the series obtained in  \cite{kg}, while the
bisphaleron branch bifurcates with the sphaleron branch at 
a value of $\beta_{BI}^2$ depending on the Higgs mass.

The model, the notations and the ansatz are given in Sect. 2,
 we discuss some special limiting solutions in Sect. 3 and present and discuss
our numerical results in Sect. 4. Finally, in Sect. 5 we comment 
on the stability of the solutions.

\section{SU(2) Born-Infeld-Yang-Mills-Higgs (BIYMH) theory}
\par Neglecting the $U(1)$ part (for technical reasons),
we consider the Lagrangian of the SU(2) part of the  electroweak model
in which the term of the Lagrangian containing the field strength
of the Yang-Mills (YM) fields is replaced by the corresponding Born-Infeld
(BI) term.
The Lagrangian density reads
\begin{equation}
{\cal L } = \beta_{BI}^2(1-R) +D_{\mu}\phi^+D^{\mu}\phi-{\lambda\over 4} (\phi^+\phi-{v^2\over 2})^2
\end{equation}
\begin{equation}
\label{BI}
R = (1+{1\over{2\beta_{BI}^2}}F^{a}_{\mu\nu}F^{a\mu\nu}-{1\over{16\beta_{BI}^4}}(F^a_{\mu\nu}\tilde F^{a\mu\nu})^2)^{1\over 2}
\end{equation}
where $\beta_{BI}$ is the coupling of the BI term and has dimension $L^{-2}$.
In the limit $\beta_{BI}^2\rightarrow \infty$, the standard $SU(2)$ electroweak Lagrangian is recovered.
We study classical, spherically symmetric, static solutions with finite
energy $E(\beta_{BI}^2, v^2,\lambda)$. A suitable rescaling
of the space variable $\vec x$ leads to a rescaling of
the energy as follows :
\begin{eqnarray}
\label{sca1}
E(\beta_{BI}^2,v^2,\lambda) &=& vE({\beta_{BI}^2\over {v^4}}, 1, \lambda)\\
\label{sca2}
&=& \sqrt{\beta_{BI}} E(1,{v^2\over {\beta_{BI}}},\lambda)
\end{eqnarray}

We are using the standard spherically symmetric ansatz of \cite{bk,yaffe}
\begin{eqnarray}
A^a_0 &=& 0 \nonumber\\ 
A_i^a &=& \frac{1-f_A(r)}{gr} \epsilon _{aij} \hat x_j
       + \frac{f_B(r)}{gr} (\delta_{ia} - \hat x_i \hat x_a)
       + \frac{f_C(r)}{gr} \hat x_i \hat x_a  \nonumber\\
\phi &=&
\frac{v}{\sqrt 2} \bigl[ H(r) + i K(r) (\hat x^a \sigma_a) \bigr]
\left(\begin{array}{c}
0\\
1\end{array}\right)
\end{eqnarray}
where $f_A(r)$, $f_B(r)$, $f_C(r)$, $H(r)$ and  $K(r)$ are functions of the radial coordinate $r$ only
and $\hat x_a \equiv r_a/r$.
It is well known (see e.g. \cite{bk,yaffe,akiba}) that this ansatz is plagued with a residual
gauge symmetry, which  we will fix here  
by imposing the axial gauge~:
\begin{equation}
      x^i A_i = 0 \ \ \  \Rightarrow \ \ \ f_C(r) = 0 \ \ .
\end{equation}
Since we consider static solutions here and due to the choice $A^a_{0}=0$, 
the term including the dual field strength tensor in (\ref{BI}) vanishes.
The masses of the gauge boson and Higgs boson, respectively, are given
by:
\begin{equation}
M_W={1\over 2}gv \ , \ M_H=\sqrt{\lambda}v
\end{equation}
With the rescaling
\begin{equation}
x=M_Wr
\label{rescale}
\end{equation}
the energy $E$ depends only on two fundamental couplings
\begin{equation}
  \beta^2\equiv {\beta_{BI}^2\over{(M_W)^4}} = 16{\beta_{BI}^2\over{g^4v^4}}
 \ , \
\epsilon \equiv {1\over 2} ({M_H\over {M_W}})^2 =  2{\lambda\over{g^2}}
\end{equation}
and is given as 1-dimensional integral over the energy
density ${\cal E}$ (the prime denotes the derivative with respect to $x$):
\begin{equation}
E(\beta^2,\epsilon) = {4\pi\over {g^2}} M_W\int dx \ {\cal E},
\end{equation}
\begin{eqnarray}
\label{energy}
{\cal E} &=& \beta^2 x^2\left(\sqrt{1+{1\over{ \beta^2x^2}}
(2(f^{'2}_A+f^{'2}_B + ({f^2_A+f^2_B-1\over x})^2)}-1\right) \nonumber \\
&+& (K(f_A+1)+H F_B)^2 + (H(F_A-1)-K f_B)^2  +2x^2(H^{'2}+K^{'2})   \nonumber  \\
&+&  \epsilon x^2(H^2+K^2-1)^2
\end{eqnarray}

\par The corresponding Euler-Lagrange equations can be obtained in a straighforward way and
have to be solved with respect to specific boundary conditions
which are necessary for the solution to be regular and of finite energy.
In \cite{bk,yaffe} it was found that two possible sets of boundary 
conditions exist. After an algebra, it can be shown that these two
sets of conditions also hold when the Yang-Mills part of the
action is replaced by the Born-Infeld term. 
The first of these sets was used to construct the solution
of \cite{km}. Two of the radial functions vanish identically~:
$f_B(x)=H(x)=0$ and the other two have to obey
\begin{eqnarray}
f_A(0) = 1\ &,&\ f_A(\infty) = - 1\nonumber\\
K(0) = 0 \ &,&\ K(\infty) = 1
\end{eqnarray}
For the second set, all functions are non-trivial and
their boundary conditions at the origin read~:
\begin{eqnarray}
&f_A(0) = 1 \ \ \ \ , \ \ \ \ &f_B(0) = 0 \nonumber \\
&H'(0) = 0 \ \ \ \ , \ \ \ \ &K(0) = 0 
\end{eqnarray}
while in the limit $x\rightarrow \infty$, 
the functions have to approach constants in the following way
\begin{eqnarray}
 &{\rm lim}_{x \rightarrow \infty} (f_A(x) + i f_B(x)) &= 
{\rm exp} (2 i \pi q) \nonumber  \\
 &{\rm lim}_{x \rightarrow \infty} (H(x) + i K(x)) &= 
{\rm exp} ( i \pi q)
\end{eqnarray}
The solutions of this second type are thus characterized by a real 
constant $q \in [0,1[$. This parameter
has to be determined numerically and
depends on $\epsilon$ and $\beta$.

\section{Limiting Solutions}
Two parameter limits of the model studied here are of special interest and have been studied previously:
\begin{description}
\item [a)] $\beta_{BI}^2=\infty$.\\
In this limit, the standard $SU(2)$ electroweak Lagrangian is obtained. 
Classical, finite energy
solutions were found for the two different sets
of boundary conditions discussed above.
The case in which $f_{B}$ and $H$ vanish is 
the sphaleron discovered by  Klinkhamer and Manton \cite{km}.
It has energy of the form
\begin{equation}
E_{sp} = {M_W\over{\alpha_W}}\ I(\epsilon)\quad , \quad \alpha_W = {4\pi\over
{g^2}}
\end{equation}
where $I(\epsilon)$ has to be determined numerically. It was found that \cite{km,bk1}
\begin{equation}
I(0) \simeq 3.04\ ,\ I(0.5) \simeq 3.64\ ,\ I(\infty)\simeq 5.41
\end{equation}
For sufficiently large values of $\epsilon$ (i. e.  $\epsilon > \epsilon_{cr}
\approx 72$)
solutions of the second type appear \cite{bk,yaffe}~: the bisphalerons 
which occur as a pair of solutions related by a parity operation.
Their classical energy is lower than the energy of the sphaleron (e.g.
$I_{bi}(\epsilon = \infty) \approx 5.07$). In the limit $\epsilon \rightarrow 
\epsilon_{cr}$ the functions describing the bisphaleron converge in an uniform way
to the functions of the corresponding sphaleron.
\item [b)] $v=0$.\\
In this case the Higgs field decouples and we are left with the pure SU(2)
Born-Infeld theory which was studied recently in \cite{kg}. While
the pure YM system does not admit finite energy solutions, the presence of the
BI term breaks the scale invariance and finite energy solutions,
so-called "glueballs" exist.
These are indexed by the number $n$ of nodes of the function $f_A^{(n)}(r)$. The first element
of the sequence, $f_A^{(1)}(r)$, has exactly one node and with the convention adopted
in  \cite{kg} the energy (in units of $4\pi$) is given by
\begin{equation}
E_{KG} =  E(1,0,0) \equiv  M
\end{equation}
It was found \cite{kg} that for the first solution of the sequence $M\simeq
1.13559$.
\end{description}
\section{Numerical results}
\par We have studied numerically the classical equations for several values
of the parameters $\beta_{BI}^2,v,\lambda$. By taking advantage of the scaling
(\ref{sca1}), (\ref{sca2}) we fixed $v=1$ without loosing generality. These solutions are therefore essentially characterized by the classical energy
$E( \beta_{BI}^2, 1,\lambda)=E( \beta^2,\epsilon)$.
\subsection{BI-sphaleron}
Starting from the sphaleron solution ($ \beta^2=\infty$) we
constructed a branch of smoothly deformed solutions up to a critical
value $ \beta^2= \beta^2_c$ along which the energy slightly decreases.
$ \beta^2_c$ depends on $\epsilon$ and is increasing for
increasing $\epsilon$ which is demonstrated in the following table:

\begin{table}[h]
\begin{center}
\begin{tabular}{|ccccccc|}
\hline
$\epsilon$ &0.5 &1 &2 &10 &100 &200\\
\hline
$ \beta^2_c$ &26.5 &37.4 &39.7 &65.0 &110. &121.\\
\hline
\end{tabular}
\end{center}
\end{table}

No solution seems to exist
for $ \beta^2 < \beta^2_c$, but we
obtain strong numerical evidence that a
second branch of solution exist for $ \beta^2 \in [\infty,  \beta^2_c]$. For
fixed $ \beta^2$, the energy of the solution of this second branch is
higher than the energy of the one of the first branch.
 We therefore refer to these branches as to the lower
and upper branch, respectively. For the upper branch we find
\begin{equation}
\lim_{ \beta^2\rightarrow \infty} E( \beta^2,\epsilon) = \infty
\end{equation}
This can be understood as follows: on the lower branch,
$ \beta^2\rightarrow \infty$ corresponds to $v$ fixed and non-equal zero
with $\beta_{BI}^2\rightarrow \infty$, while on the upper branch this limit corresponds
to $\beta_{BI}^2$ fixed and $v\rightarrow 0$. Because of the special choice of
the coordinates (\ref{rescale}), the solution on the upper branch
shrinks to zero (see also Fig. 4), while its energy tends to infinity
for $ \beta^2\rightarrow \infty$.
With an appropriate rescaling
$E( \beta^2, \epsilon)$ $\rightarrow$
 $(\beta)^{-{1\over 2}} E(
\beta^2, \epsilon)$
the mass of the $n=1$ Kerner-Galtsov solution is recovered
\begin{equation}
M = \lim_{ \beta\rightarrow \infty}  \beta^{-{1\over 2}}
E( \beta^2, \epsilon)
\end{equation}
For $\epsilon=0.5$, the energy of the solutions on both branches
before the rescaling is illustrated in Fig. 1, the energy after
the rescaling in Fig. 2.

In Fig. 3 we show the value $x_0$ of the radial coordinate $x$, for which
$f_A^{(1)}$ attains its node. Clearly, this value tends to zero for
$ \beta^2\rightarrow\infty$ on the upper branch. As already mentioned
this is due to the fact that the rescaled variable $x$ shrinks to zero
for $v\rightarrow 0$.
Also shown in Fig. 3 is the contribution $E_{Higgs}$ of the Higgs field energy to the total energy
of the solution. While on the lower branch it stays finite for
all values of $\beta^2$, it tends to zero on the upper branch for $\beta^2\rightarrow\infty$ .
This supports the interpretation that on the upper branch
the Higgs field is trivial
for $ \beta^2=\infty$.

Finally, we demonstrate the convergence of the $n=1$-solution to the corresponding
Kerner-Gal'tsov (KG) solution for $\epsilon=0.5$ in Fig. 4. The profiles
of the gauge field and Higgs field functions are shown on the lower branch
for $\beta^2=100$ and a value $\beta^2=27$
close to the critical $\beta^2_{c}$. Clearly, the functions
tend to that of the KG solution on the upper branch for increasing
$\beta^2$. In the limit $\beta^2\rightarrow\infty$, $f_A$
tends to the gauge field function of the KG solution, while the Higgs
field function $K$ tends to zero on the full intervall $[0:\infty[$.

\subsection{BI-bisphaleron}
For sufficiently high value of the Higgs-boson mass (i.e. for $\epsilon > 72$)
bisphaleron solutions can be constructed.
As in the case of the sphaleron, bisphalerons are smoothly deformed by the
Born-Infeld parameter. By studying these solutions in detail
for $\epsilon=100$
and $\epsilon=200$ and varying $\beta^2$, we observed that like the BI-sphalerons
the deformed bisphalerons exist up to a critical value  $\beta^2 =
\tilde{\beta}_{c}^2$.
For $\epsilon =100$, the BI-bisphaleron branch merges into the
lower BI-sphaleron branch at $\beta^2 = \tilde{\beta}_{c}^2 $. For still
higher values of $\epsilon$ 
 the pattern is slightly different.
Another branch of BI-bisphaleron exist on the interval
$\beta \in [ \tilde{\beta}_{c}^2, \hat{\beta}^2]$ and the solutions on this 
branch merge into the upper BI-sphaleron branch in the limit
$\beta^2 \rightarrow \hat{\beta}^2$. In both cases, all radial functions
characterizing the BI-bisphaleron converge in an uniform way
to the radial functions associated with the corresponding BI-sphaleron.
These various phenomena are illustrated in Fig. 5 where the value
$H(0)\propto \vert \phi(0) \vert$ 
is shown as function
of $1/\beta^2$ for $\epsilon=100$ and $\epsilon=200$.
For $\epsilon =100$, this quantity clearly tends to zero for $
\tilde{\beta}_{c}^2 = \hat{\beta}^2 \approx 113.6$. For
$\epsilon =200$, we find that $\tilde{\beta}_{c}^2 \approx 108.8$
(with $H(0)\neq 0$), while the bisphaleron bifurcates with the sphaleron
(with $H(0)$ tending to zero) at
$\hat{\beta}^2 \approx 164$.
We further observed that for fixed $\beta^2$ and for
both lower and upper  branches, the energy of the BI-bisphaleron 
is close, although  lower, than the energy of the BI-sphaleron. 
In the limit $\beta^2 \rightarrow \hat{\beta}^2$, the two
upper branches (if any) merge into a single one corresponding
to the BI-sphaleron upper branch.

\section{Stability}

Let us finally discuss the (in)stability of our solutions.
When two branches of solutions do exist terminating into
a catastrophe spike, like the ones shown in Figs.~1 and 2,
 it is widely believed
that the number of negative modes is constant along each
branch and that the number of negative modes on the branch with
higher energy exceeds the number of negative
modes on the branch with lower energy by one unit.
The reasoning is based on catastrophe theory \cite{kus}
and was demonstrated to hold in the context of classical
solutions in various models (e.g. \cite{bks,bk2}).

For $\epsilon < 72$, the lower branch corresponds to the 
deformed electroweak sphaleron and 
thus possesses a single direction of instability. 
Therefore, the solutions on the lowest branch likely have one
direction of instability while the solutions on the upper
branch possess two. This is not in contradiction with the result
of \cite{kg} because these solutions, when embedded
into a field theory with extra fields (here the Higgs field),
likely acquire extra unstable modes due to the supplementary
degrees of freedom.

For the values of $\epsilon > 72$
we have considered here, the bisphaleron possesses one direction of 
instability on the lower branch while the sphaleron has two.
On its upper branch (i.e. for $\hat{\beta}^2 \geq \beta^2 \geq \tilde{\beta}_c^2$)
the BI-bisphaleron (resp. BI-sphaleron) possesses two (resp. three)
directions of instability. For $\beta^2 \geq \hat{\beta}^2$ only the upper
branch of the BI-sphaleron exists and it likely possesses 
two unstable modes, similarly to the case $\epsilon < 72$.

Finally, for $\epsilon > 9500$ the situation is still more involved 
\cite{bk,yaffe}. Nevertheless,
the BI-bisphaleron possesses a single 
direction of instability on its lower branch
and the above conclusions should apply, too.

\section{Summary}
We have studied the Born-Infeld deformation of
the SU(2) electroweak sphaleron and bisphaleron.
We find that the solutions exist up to a critical value of the
Born-Infeld coupling $\beta_{c}^2$ (resp. $\tilde{\beta}_{c}^2$) 
which depends on the Higgs mass parameter.

The fact that classical solutions of Born-Infeld like gauge
field theories cease to exist at a critical value of the parameter $\beta^2$
was observed in several models with Abelian \cite{sha1} and
non-Abelian \cite{sha2,sha3} gauge groups. Especially
for the case of BI-vortices, it was shown recently
\cite{bm} that the magnetic field $B(0)$ of the solution
becomes infinite when the critical value $\beta_c^2$ is approached. 

The situation here is slightly different. Indeed, the solutions
exist only for $\beta^2 > \beta_{c}^2$, but
we constructed numerically a second branch of
solutions existing up to $\beta^2=\infty$. There the
solution tends to the first solution
of the KG series and - after an appropriate rescaling - the energy on this
upper branch reaches the mass of the $n=1$ KG solution in the limit
$\beta^2\rightarrow\infty$.
When BI-bisphaleron are present, the pattern is qualitatively
similar: the BI-bisphaleron branch bifurcates from one of the
BI-sphaleron branches (upper or lower, depending of the value of $\epsilon$).
No BI-bisphalerons  exist as an upper branch for sufficiently 
high values of $\beta^2$.

Sphaleron solutions  can also be studied in models where the Born-Infeld
term is incorporated by using a symmetrized trace.  The result of
\cite{dyadi} suggests though, that no
qualitative changes of the critical pattern should occur.

%%%%\end{document}

\newpage
\begin{fixy}{-1}
\begin{figure}
\mbox{\epsffile{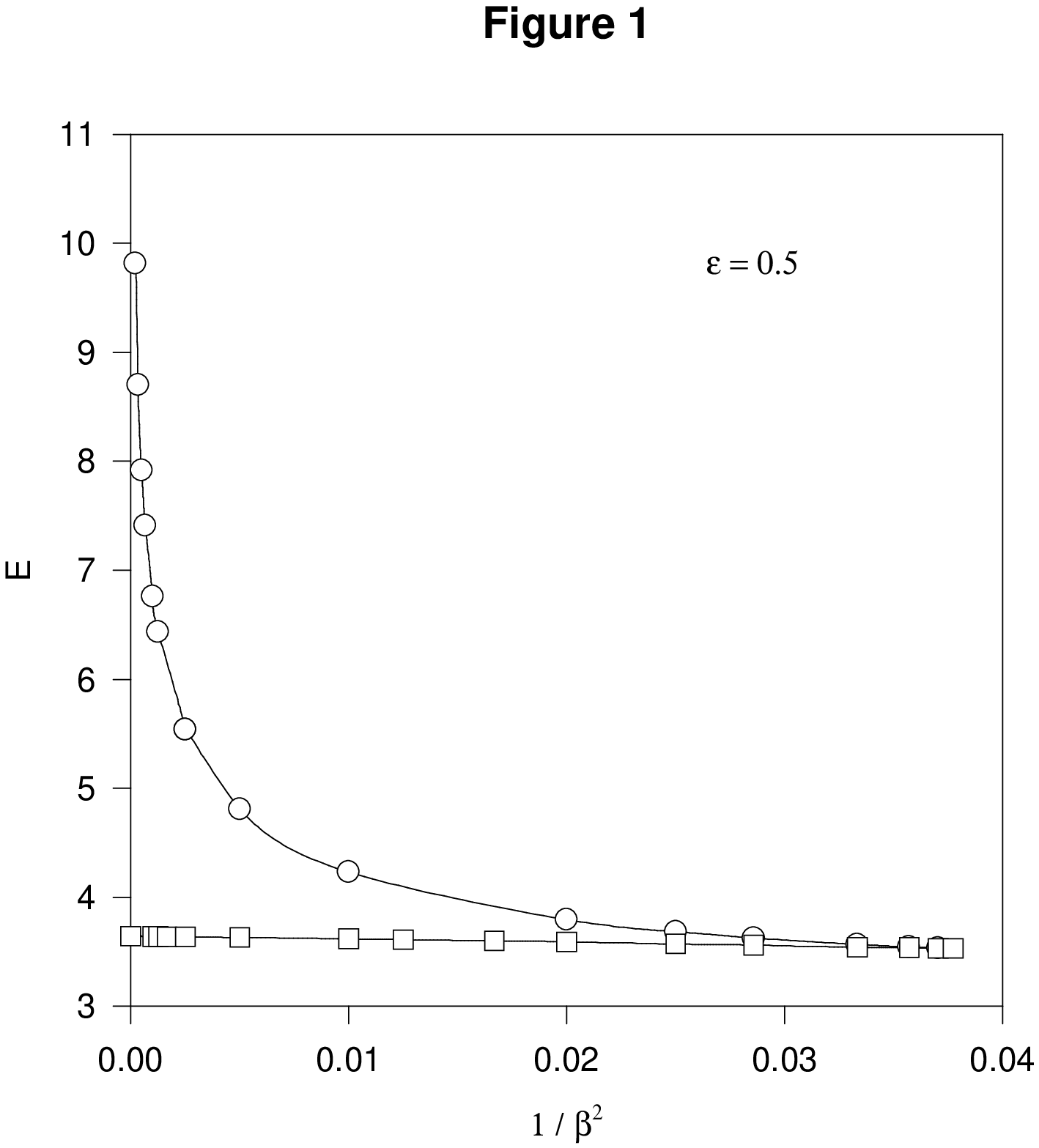}}
\caption{The energy $E$ is shown
as a function of $ 1/\beta^2$ for $\epsilon=0.5$.}
\end{figure}
\end{fixy}

\newpage
\begin{fixy}{-1}
\begin{figure}
\mbox{\epsffile{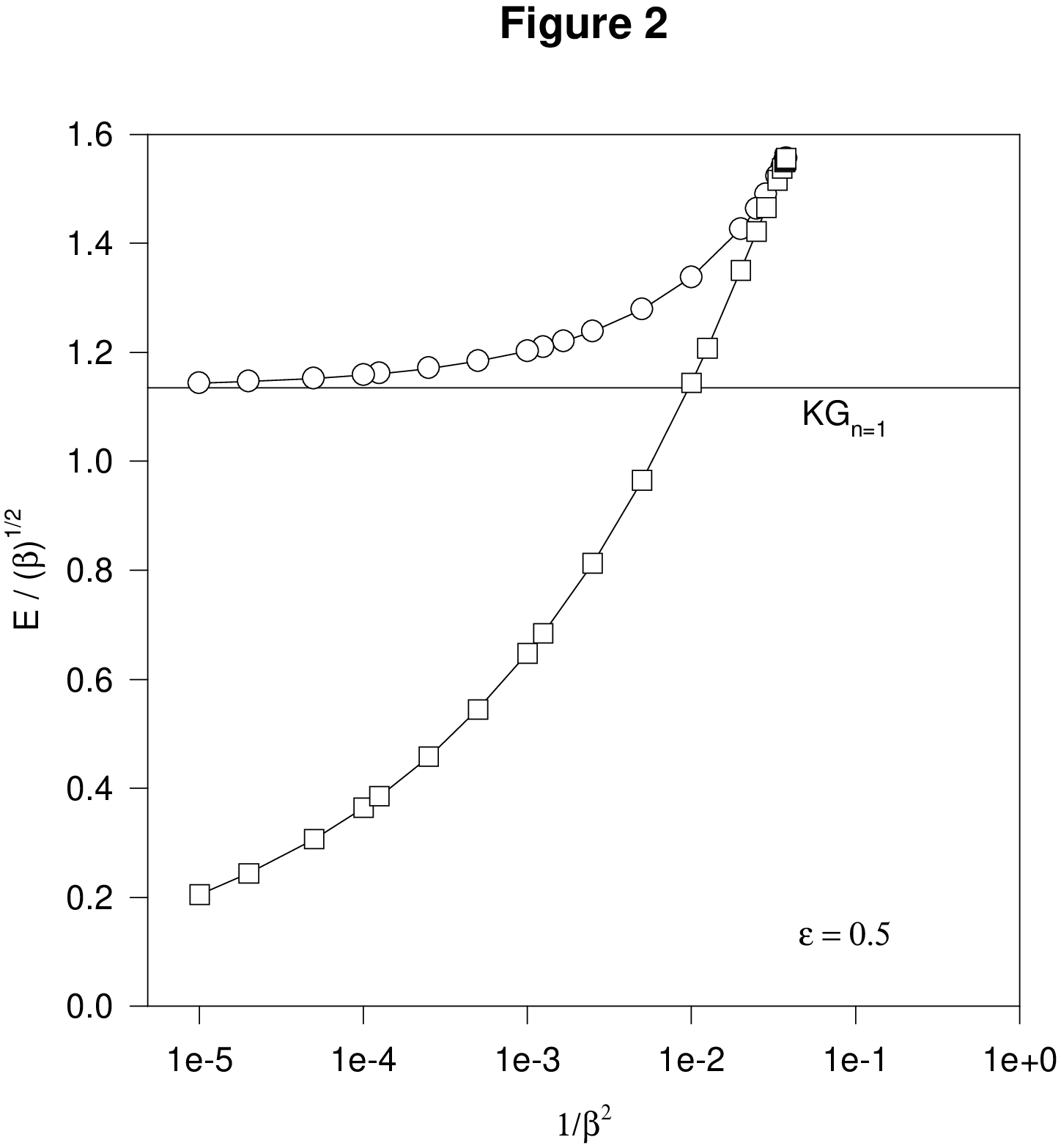}}
\caption{The ratio $E(1/\beta^2)/\beta^{1\over 2}$ is shown as a function of
 $1/ \beta^2$ for $\epsilon = 0.5$. The horizontal solid line
represents the energy of the $n=1$ KG solution in units of $4\pi$ with $M\simeq 1.135$.}
\end{figure}
\end{fixy}

\newpage
\begin{fixy}{-1}
\begin{figure}
\mbox{\epsffile{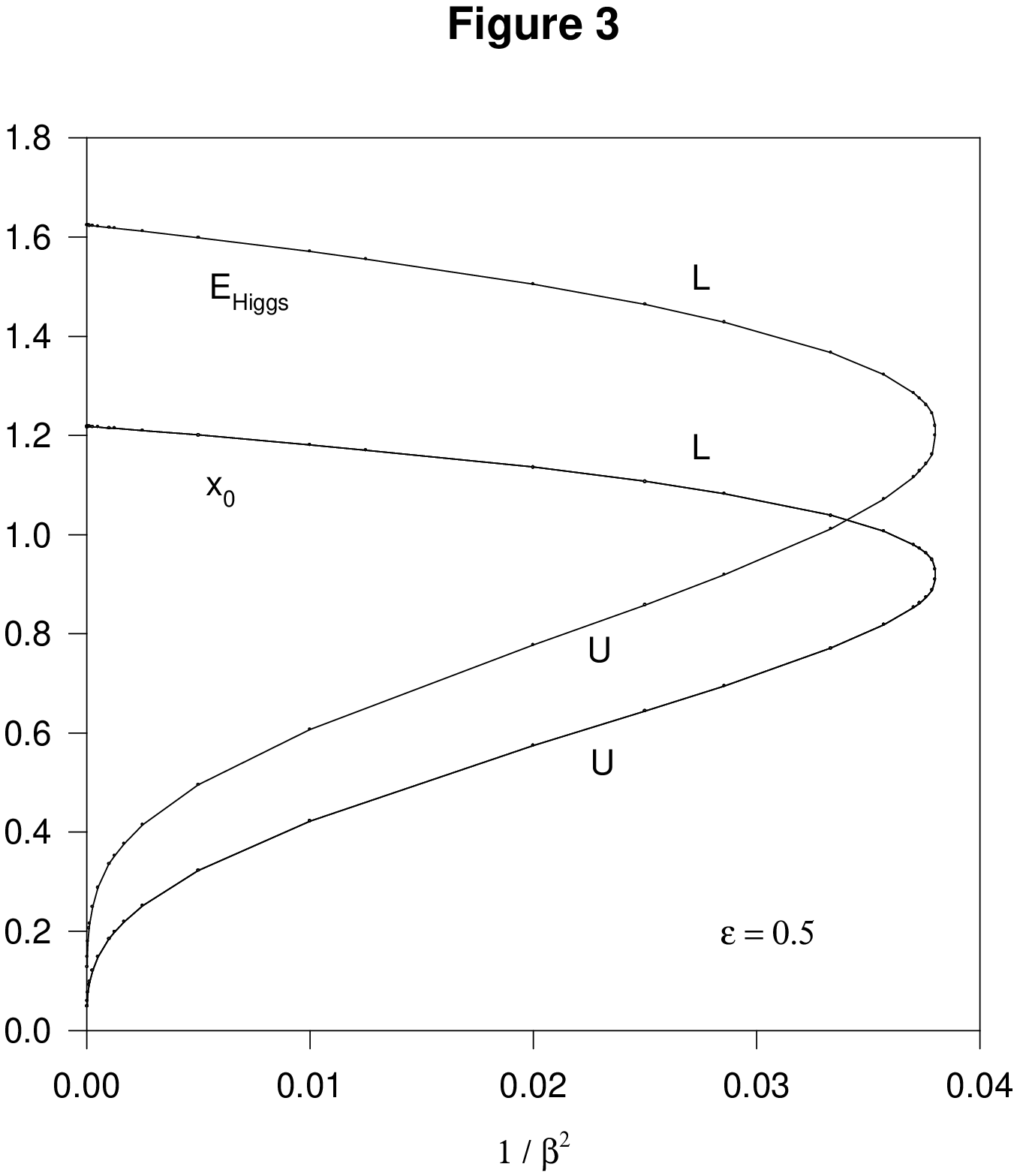}}
\caption{The value $x_0$ for which $f_A^{(1)}(x_0)=0$ as well as
the contribution $E_{Higgs}$ of the Higgs field
energy
to the total energy of the
solution are plotted as functions of $1/\beta^2$ for $\epsilon=0.5$
. The labels $L$ and $U$,
respectively, refer to the lower and upper branch.}
\end{figure}
\end{fixy}

\newpage
\begin{fixy}{-1}
\begin{figure}
\mbox{\epsffile{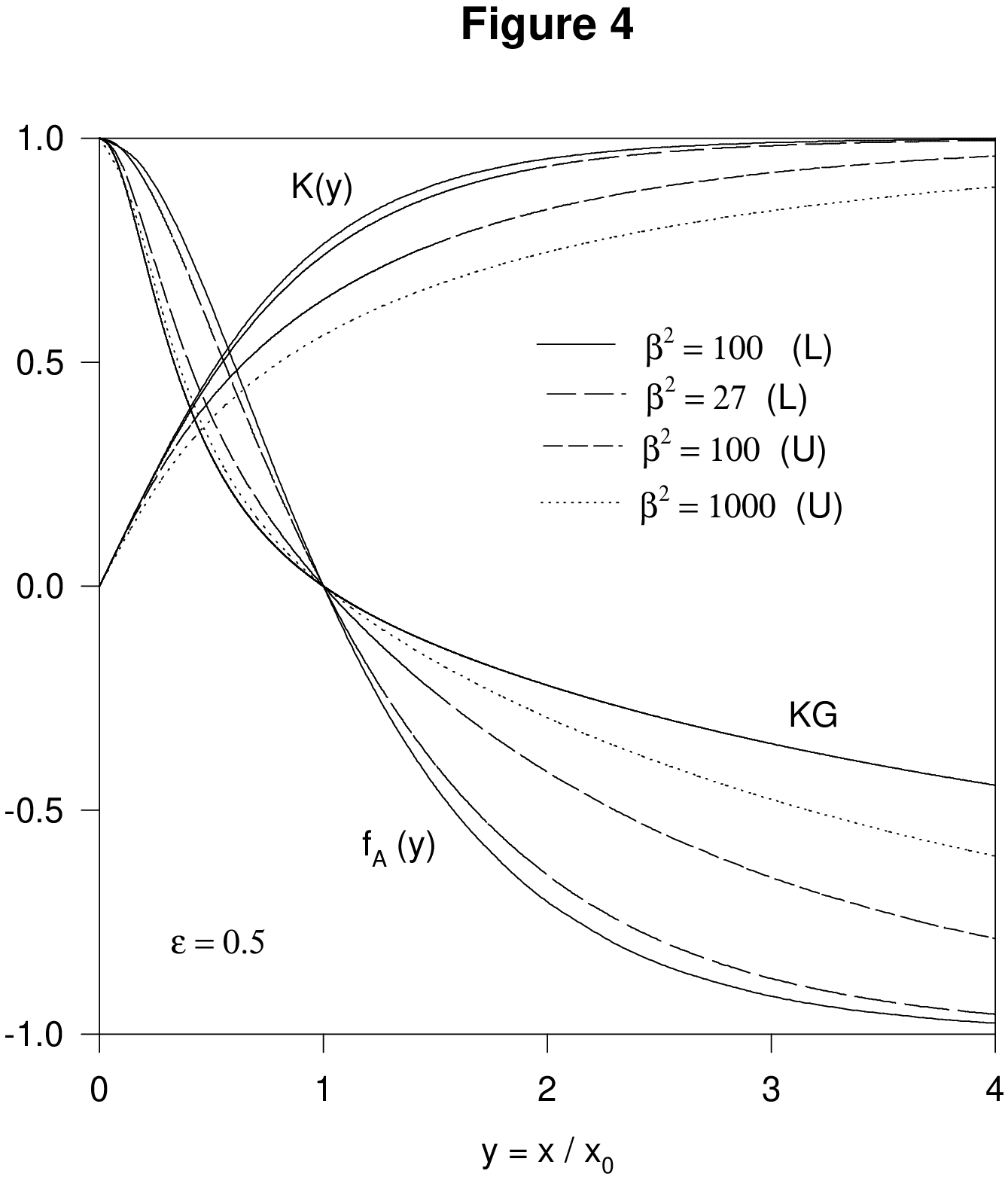}}
\caption{The profiles of the solutions $(f_A^{(1)}, K)$ as function of
$y=x/x_0$ are shown for several values of $\beta^2$ and $\epsilon=0.5$.  For comparison, the profile
of the function $f_A^{(1)}$ of the $n=1$ KG solution is also shown. $K(y)\equiv 0$ for the KG solution.}
\end{figure}
\end{fixy}

\newpage
\begin{fixy}{-1}
\begin{figure}
\mbox{\epsffile{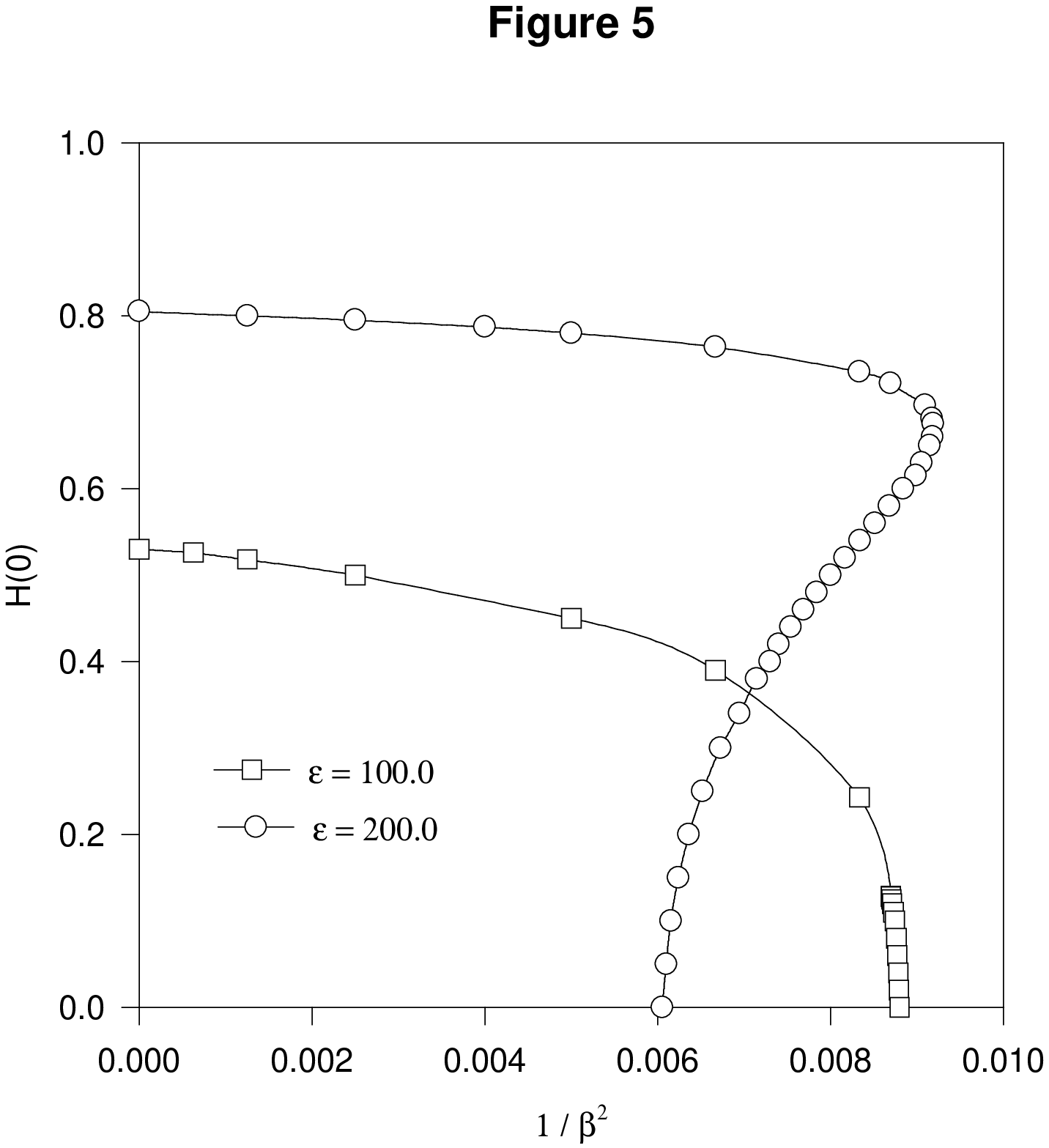}}
\caption{The  value $H(0)$ $\propto$ $\vert \phi(0) \vert$ characterizing
the bisphaleron solution is shown as a function of $1/\beta^2$ for
two values of $\epsilon$.}
\end{figure}
\end{fixy}

\end{document}